\documentclass[sigconf]{acmart}

\usepackage{enumitem}
\setlist[enumerate,itemize]{leftmargin=*, topsep=0pt}
\usepackage{booktabs}
\usepackage{graphicx}
\usepackage{caption}
\usepackage{subcaption}

\usepackage{booktabs}
\usepackage{multirow,multicol}
\usepackage[utf8]{inputenc}
\usepackage{algorithm}
\usepackage{algpseudocode}
\usepackage{amsmath}
\usepackage{pifont}

\usepackage{changes}

\usepackage{amsmath, amssymb}
\usepackage[framemethod=tikz]{mdframed}

\mdfdefinestyle{customstyle}{
  linecolor=gray!100,
  linewidth=3pt,
  innerleftmargin=3pt,
  topline=false,
  rightline=false,
  bottomline=false,
  leftline=true,
  innerrightmargin=3pt,
  innertopmargin=3pt,
  innerbottommargin=3pt,
  backgroundcolor=gray!15
}

\newenvironment{custommdframed}
  {\begin{mdframed}[style=customstyle]}
  {\end{mdframed}}

\newcommand{\ourmodel}{\textsc{SecureReviewer}}

\begin{document}

\title{\ourmodel{}: Enhancing Large Language Models for Secure Code Review through Secure-aware Fine-tuning}

\author{Fang Liu$^{1}$, Simiao Liu$^{1}$, Yinghao Zhu$^{1}$, Xiaoli Lian$^{1}$, Li Zhang$^1$$^\ast$}
\thanks{$^{\ast}$Corresponding author.}
\affiliation{%
\institution{$^1$State Key Laboratory of Complex \& Critical Software Environment, School of Computer Science and Engineering, \\ Beihang University, China}
  \country{}
}

\email{{fangliu,buaalsm,zhuyinghao,lianxiaoli,lily}@buaa.edu.cn}

\begin{abstract}
Identifying and addressing security issues during the early phase of the development lifecycle is critical for mitigating the long-term negative impacts on software systems. Code review serves as an effective practice that enables developers to check their teammates' code before integration into the codebase. To streamline the generation of review comments, various automated code review approaches have been proposed, where Large Language Model (LLM)-based methods have significantly advanced the capabilities of automated review generation. However, existing models primarily focus on general-purpose code review, their effectiveness in identifying and addressing security-related issues remains underexplored. Moreover, adapting existing code review approaches to target security issues faces substantial challenges, including data scarcity and inadequate evaluation metrics. To address these limitations, we propose \ourmodel{}, a new approach designed for enhancing LLMs' ability to identify and resolve security-related issues during code review. Specifically, we first construct a dataset tailored for training and evaluating secure code review capabilities. Leveraging this dataset, we fine-tune LLMs to generate code review comments that can effectively identify security issues and provide fix suggestions with our proposed secure-aware fine-tuning strategy. To mitigate hallucination in LLMs and enhance the reliability of their outputs, we integrate the Retrieval-Augmented Generation (RAG) technique, which grounds the generated comments in domain-specific security knowledge. Additionally, we introduce SecureBLEU, a new evaluation metric designed to assess the effectiveness of review comments in addressing security issues. Experimental results demonstrate that \ourmodel{} outperforms state-of-the-art baselines in both security issue detection accuracy and the overall quality and practical utility of generated review comments. Our code and data are available at \url{https://github.com/SIMIAO515/SecureReviewer}.
\end{abstract}



\copyrightyear{2026}
\acmYear{2026}
\setcopyright{cc}
\setcctype{by}
\acmConference[ICSE '26]{2026 IEEE/ACM 48th International Conference on Software Engineering}{April 12--18, 2026}{Rio de Janeiro, Brazil}
\acmBooktitle{2026 IEEE/ACM 48th International Conference on Software Engineering (ICSE '26), April 12--18, 2026, Rio de Janeiro, Brazil}
\acmPrice{}
\acmDOI{10.1145/3744916.3773191}
\acmISBN{979-8-4007-2025-3/26/04}

\maketitle
\section{Introduction}
As software systems play increasingly critical roles in society, security vulnerabilities can have profound consequences for businesses and individuals~\cite{mcgraw2004software,bavota2015four}. To mitigate these risks, modern software development adopts proactive "shift-left" practices~\cite{ichu2011role,dawoud2024better} that integrate security testing into earlier development phases. Code review serves as a key preventive measure in this paradigm, where developers submit code changes for systematic evaluation to identify and address issues before codebase integration~\cite{fagan2002history,tufano2021towards}. For example, Heartbleed (CVE-2014-0160)~\cite{cve20140160}, a famous OpenSSL vulnerability from improper input validation, could have been prevented through effective code review~\cite{durumeric2014matter}. Furthermore,~\citet{bavota2015four} find that unreviewed commits are more than twice as likely to introduce bugs and are less readable than reviewed ones.

There are several empirical studies that explore the role of code review in finding and mitigating security issues~\cite{yu2023security,charoenwet2023complementing,yu2024insight}. While these studies identify key challenges and limitations in practice and provide valuable recommendations and insights for improving secure code reviews, they do not offer automated solutions to systematically address these issues.
To efficiently generate review comments and reduce reliance on manual effort, recent years have seen the emergence of automated code review approaches, leveraging the rapid advancements in deep learning technologies~\cite{gupta2018intelligent,shi2019automatic,tufano2022using}.

For example,~\citet{gupta2018intelligent} introduce an LSTM-based model designed to analyze the relationships between code changes and review comments, and recommends review comments automatically based on existing code reviews and code changes. 
Building on advances in Transformer~\cite{vaswani2017attention} architectures and pre-trained models~\cite{raffel2020t5,feng2020codebert,wang2021codet5}, researchers have developed code review systems through two primary approaches: pre-training models on code review-specific tasks~\cite{li2022codereviewer} or fine-tuning large language models (LLMs) for code review applications~\cite{lu2023llama,yu2024fine}. These approaches, particularly LLM-based methods, have significantly advanced the capabilities of automated review comment generation, pushing the boundaries of what is achievable in this domain and creating new opportunities for secure code review practices. However, existing models focus primarily on general-purpose code review, and their effectiveness in identifying and addressing security-related issues remains unexplored~\cite{basic2024large,wang2024your}. Furthermore, adapting current code review approaches to specifically target security issues faces the following challenges: 

\noindent\textbf{\ding{182} Noisy Code Review Dataset:} Existing commonly used code review datasets~\cite{tufano2022using,li2022codereviewer} are primarily collected from generative-purpose review comment in open-source projects, where many of the comments may lack substantive content are often unrelated to identifying actual issues~\cite{yu2024fine}. For instance, the comments often contain non-informative content such as mentions of authors' names or generic statements like ``\textit{Looks good to me}'' or ``\textit{Why do we need this?}'', rather than pinpointing specific issues.
Moreover, there is a scarcity of high-quality, security-focused datasets specifically constructed for training and evaluating code review models. 

\begin{figure}[t]
    \centering
    \setlength{\abovecaptionskip}{0.1cm}
    \includegraphics[width=0.9\linewidth]{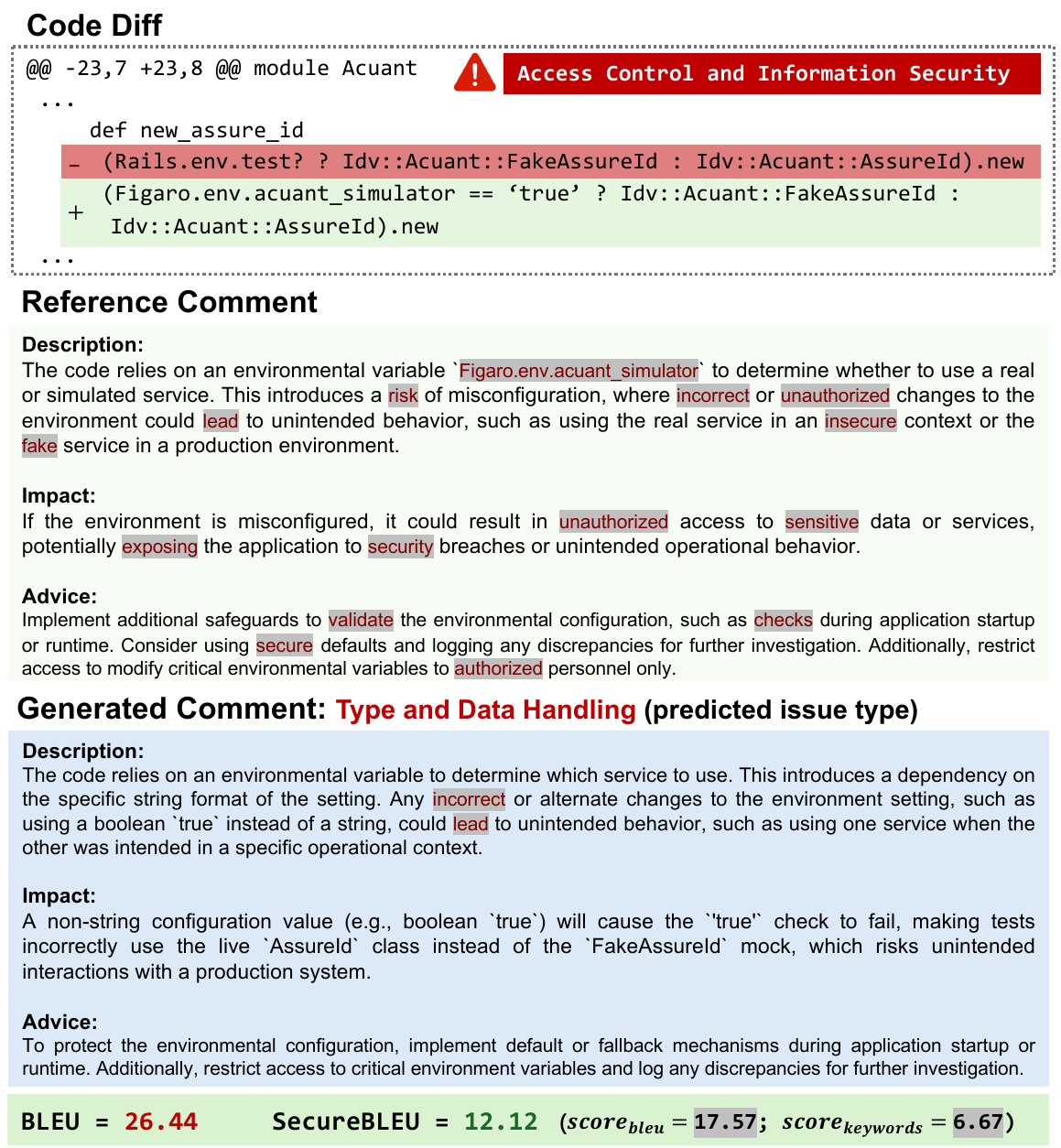}
    \caption{A code review comment with its SecureBLEU score.}
    \label{fig:bleu_exp_new}
    \vspace{-0.5cm}
\end{figure}

\noindent\textbf{\ding{183} Inadequate Metric:} BLEU score~\cite{papineni2002bleu} is widely adopted in assessing the quality of the review comment by measuring the n-gram overlap between the ground truth and predicted comment~\cite{li2022codereviewer,lu2023llama}, which fails to fully assess the effectiveness of the comments in detecting and resolving security issues. As illustrated in Figure~\ref{fig:bleu_exp_new}, even though the generated comment incorrectly classified the security issue from ``Access Control and Information Security'' to ``Type and Data Handling'' in the code diff, the BLEU score remains relatively high due to the surface-level linguistic similarity between the generated comment and the ground truth.

We propose \ourmodel{} to enhance an LLM's security code review capabilities. Our approach first involves an automated data workflow, which integrates LLMs and heuristic rules to build a tailored dataset for training and evaluation. Leveraging this dataset, we devise a security-aware fine-tuning strategy that trains the model to generate precise comments identifying security vulnerabilities and proposing actionable fixes. To further improve comment relevance and mitigate hallucinations, we employ Retrieval-Augmented Generation (RAG)~\cite{lewis2020retrieval}, which grounds the generation process by retrieving relevant examples from a prebuilt datastore of review templates.

To assess comment quality, we introduce SecureBLEU, a novel metric designed to evaluate the effectiveness of comments in identifying and resolving security issues.
We evaluate \ourmodel{} on our constructed security dataset, performing a comprehensive comparison against state-of-the-art (SOTA) code review baselines and leading LLMs. The results demonstrate that \ourmodel{} surpasses these baselines in both security issue detection accuracy and the overall quality of generated comments. In summary, our contributions are:
\begin{itemize}
    \item We design an automated data collection and refinement pipeline to construct the dataset specially designed for training and evaluating the model's capabilities of secure code review.
    \item We propose a secure-aware fine-tuning strategy, enhancing LLM to focus on generating code review comments that can effectively identify security issues and provide fix suggestions.
    \item We design SecureBLEU, a new evaluation metric for assessing the quality of code review comments by incorporating domain-specific relevance to security.
    \item We conduct a comprehensive comparison between \ourmodel{} and state-of-the-art baselines. The evaluation results demonstrate the effectiveness and practicality of our model.
\end{itemize}

\begin{figure*}[t]
    \setlength{\abovecaptionskip}{0.1cm}
    \centering
    \includegraphics[width=0.9\linewidth]{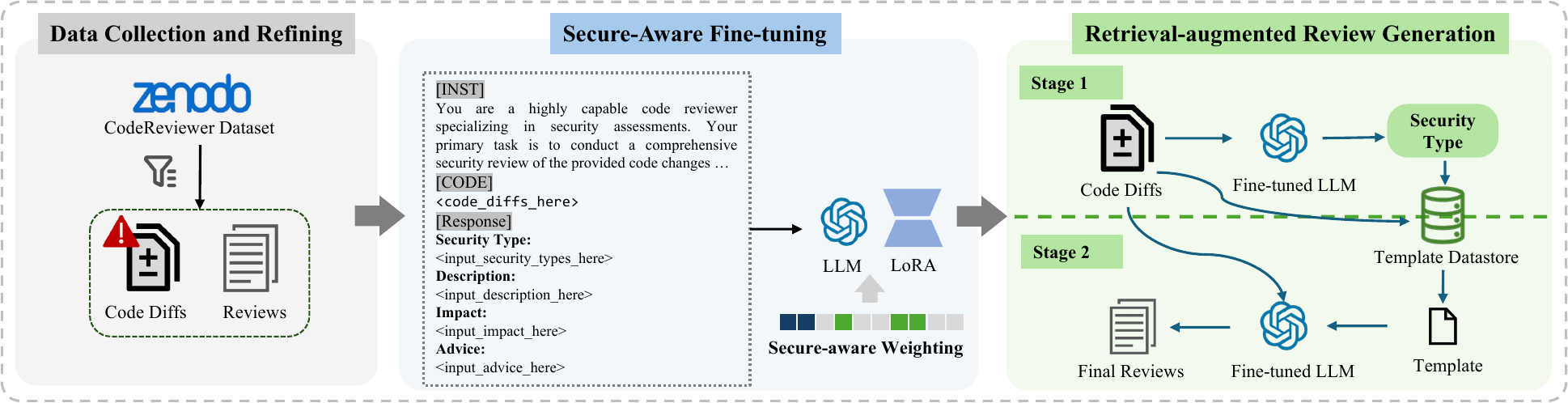}
    \caption{Overview of \ourmodel{}.}
    \label{fig:overview}
    \vspace{-0.3cm}
\end{figure*}

\section{Methodology}

Figure~\ref{fig:overview} presents the overview of \ourmodel{}. Our approach begins with the construction of a high-quality dataset to enable effective training and evaluation of the model's secure code review capabilities. Based on our dataset, we fine-tune LLM to focus on generating code review comments that can precisely identify security issues and provide fix suggestions with our proposed secure-aware fine-tuning strategy. Finally, we integrate the RAG technique to enhance the relevance and reliability of generated review comments.

\subsection{Data Collection and Refining}
\begin{figure}[t]
    \centering

    \setlength{\abovecaptionskip}{0.1cm}
    \includegraphics[width=\linewidth]{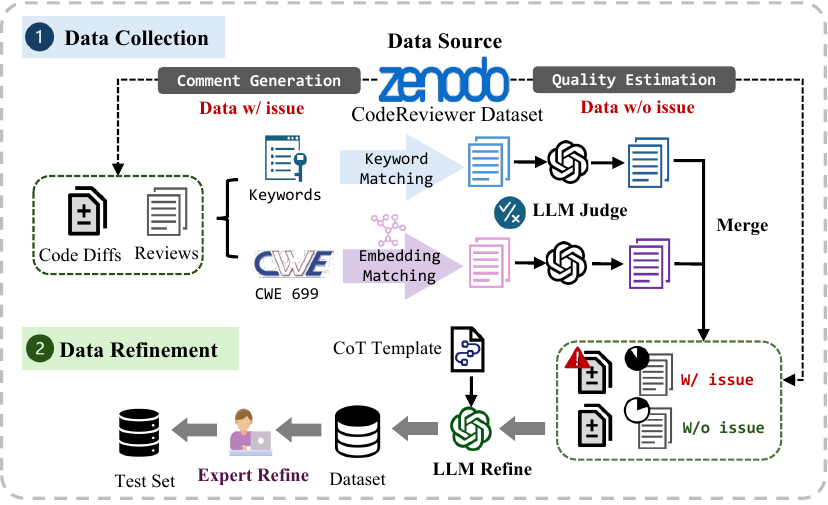}
    \caption{The process of data collection and refining.}
    \label{fig:data_collection}
    \vspace{-0.5cm}
\end{figure}

As illustrated in Figure~\ref{fig:data_collection}, we design an automated data collection and refinement pipeline, integrating both LLMs and heuristic rules, to construct the dataset for secure code review.

\subsubsection{Data Collection}
We adopt the CodeReviewer dataset~\cite{li2022codereviewer} as our primary data source, as it is a large-scale dataset curated from pull requests from well-regarded GitHub projects that includes detailed code changes ($\textit{code}_{\textit{diff}}$), commit logs, review comments ($R$), covering nine programming languages. It provides a comprehensive representation of real-world code review practices, making it well-suited for training and evaluating our secure code review model. Specifically, the dataset consists of three sub-datasets corresponding to three downstream tasks: code change quality estimation, review comment generation, and code refinement. Both the code change quality estimation and review comment generation datasets are utilized to construct our dataset. 

Due to a large number of code review comments (about 138K comments), it is not feasible for us to manually identify comments related to security issues.
To address this, we combine keyword matching and semantic embedding matching~\cite{wu2018word} methods to capture both explicit mentions of security issues and implicit references to secure coding practices.

\noindent\textbf{\textit{Keyword Matching.}} To extract security weaknesses, we employ keyword matching using a set from~\citet{yu2023security}. From the original 122 keywords spanning 15 security defect types, we exclude the ``common keywords'' category to reduce noise. Each defect type is mapped to a Common Weakness Enumeration (CWE)\footnote{\url{https://cwe.mitre.org/}}. After text normalization (lowercasing, stemming, and punctuation removal), this initial filtering yields 10,840 candidate comments.
To ensure high precision, we further refine this set using GPT-4o~\cite{achiam2023gpt4} as an LLM Judge~\cite{zheng2023judging,DevBench}. For each candidate, the LLM Judge receives the code change, the full review comment, and the matched security type. It then performs a binary classification on whether the comment accurately reflects the security issue, resulting in a final, high-quality dataset of 1,995 security-tagged review comments.

\noindent\textbf{\textit{Embedding Matching.}}
To identify security-related comments lacking explicit keywords, we employ an embedding-based matching approach. We generate vector representations for review comments and Common Weakness Enumeration (CWE) descriptions using SO\_word2vec~\cite{efstathiou2018word}, a model tailored for the software engineering domain.
As our semantic anchors, we leverage descriptions from CWE-699~\cite{cwe699}, a structured vulnerability classification focused on the software development lifecycle. This classification organizes over 400 individual weaknesses into 40 major categories and notably provides the keyword groups utilized in our preceding keyword matching step.
The process involves pre-processing both text sources (e.g., removing stop words and normalizing) and then computing the cosine similarity~\cite{singhal2001modern} between comment and CWE vectors. We retain pairs exceeding a 70\% similarity threshold, which was empirically chosen over 65\% and 75\% to best balance match quality and quantity. This candidate set is then filtered using the identical LLM Judge process from our keyword matching stage, ultimately yielding 2,771 security-tagged review comments.

\begin{table*}[t]
\centering
\footnotesize
\setlength{\abovecaptionskip}{0.1cm}
\caption{Statistics of our dataset.
}
\label{tab:final_categories}
\begin{tabular}{p{0.2\textwidth} | p{0.25\textwidth} | p{0.25\textwidth}| p{0.05\textwidth}| p{0.1\textwidth}}
\toprule
\textbf{Security Type} & \textbf{Keyword} & \textbf{CWE IDs} & \textbf{Count} & \textbf{Prop. (\%)} \\
\midrule
Exception Handling & Crash & CWE-389, CWE-429, CWE-1228  &532 & 11.38\\
\midrule
Concurrency & Race Condition, Deadlock & CWE-557, CWE-387 &412&8.81\\
\midrule
Input Validation & SQL Injection, Format String, Command Injection & CWE-1215, CWE-133, CWE-137 &819 &17.52\\
\midrule
Access Control and Information Security & Improper Access,  Cross Site Scripting (XSS), Cross Site Request Forgery, Encryption & CWE-1211, CWE-1212, CWE-1210, CWE-255, CWE-417, CWE-310, CWE-320, CWE-1216, CWE-275, CWE-265, CWE-355, CWE-1217, CWE-199&795&17.01\\
\midrule
Resource Management & Buffer Overflow, Use After Free, Resource Leak & CWE-1218, CWE-411, CWE-465, CWE-452, CWE-1219, CWE-399&292	&6.25 \\
\midrule
State Management & Denial of Service (DoS) & CWE-1006, CWE-438, CWE-840, CWE-1226, CWE-1225, CWE-371&740&15.83 \\
\midrule
Type and Data Handling & Integer Overflow & CWE-1214, CWE-1227, CWE-569, CWE-1213, CWE-189, CWE-136, CWE-19 &499&10.68\\
\midrule
Non-Issue & - & -&585 &12.52\\
\bottomrule
\end{tabular}
\vspace{-0.3cm}
\end{table*}

\noindent\textit{\textbf{Data Combination.}}
We integrate the data gathered through keyword and embedding matching by removing duplicates, merging similar security types, and ensuring a balanced distribution across security types.
Specifically, we begin by eliminating duplicate entries, resulting in an initial dataset of 4,089 unique data instances from 4,766. Next, we consolidate semantically similar security types to reduce redundancy. 
Additionally, types with low sample counts are merged with related types to improve the overall balance of the dataset~\cite{chawla2002smote}.
Finally, we derive seven security types, as detailed in Table~\ref{tab:final_categories}. 
To ensure a realistic representation of real-world code scenarios and mitigate class imbalance, we further incorporate ``Non-Issue'' data from the ``\textit{code change quality estimation}'' task in CodeReviewer dataset as the 8-th type, where instances without any code review comments are considered as ``Non-Issue''~\cite{li2022codereviewer}, and 585 samples are selected to maintain balance with the other types (approximately 1/8 of the whole dataset).  
With the inclusion of this additional category, the final dataset comprises 4,674 entries, and the distribution of category proportions is illustrated in Table~\ref{tab:final_categories}.

\subsubsection{Data Refinement}\label{refinement}
Original review comments frequently contained ambiguous phrasing or lacked critical elements essential for comprehensive security code review. To address these limitations, we perform systematic data refinement that adapts the principles of effective code review~\cite{kononenko2016code,yu2024fine} to the security context.
Specifically, we decomposed the secure code review task into the following four sequential sub-tasks:
\begin{itemize}
    \item \textbf{Identify the Security Type}: Clearly specify the type of security issue that is being addressed.
    \item \textbf{Describe the Issue}: Provide a clear and logical description of the root cause of the identified issue.
    \item \textbf{Explain the Impact}: Analyze the potential impact of the issue, laying the foundation for proposing a solution.
    \item \textbf{Advise an Improvement}: Offer actionable and specific recommendations to resolve the issue.
\end{itemize}
We argue the security code review comment should encompass the above elements, and \textbf{we formally define the security code review comment $R$ as: $R = (ST, D, I, A)$}, where $ST$ represents the Security Type, $D$ is the issue description, $I$ denotes the impact, and $A$ provides actionable advice for resolving the issue.

Leveraging the advanced capabilities of LLMs in tasks such as code understanding~\cite{ma2024understand}, vulnerability detection~\cite{du2024vul}, and bug fixing~\cite{xia2024chatrepair}, we employ GPT-4o to automatically refine the collected review comment data using one-shot prompting guided by the aforementioned criteria, transforming raw review comments into structured, comprehensive review comment.

\noindent\textbf{Initial Data Quality Assessment:}
To validate the quality of the LLM refined data, we randomly sampled 351 data entries from the 4,089 refined pieces (achieving 95\% confidence level with 5\% confidence interval~\cite{bulpitt1987confidence}). Two domain experts, each with over 6 years of software development experience, independently evaluated these samples using aforementioned four criteria.
This initial validation required 8-10 minutes per sample for code understanding and security validation, including bidirectional verification with the original comment, totaling 98 person hours.
The experts achieved a Cohen's Kappa score of 0.74, indicating substantial inter-rater agreement, with disagreements resolved through discussion.
This validation confirmed that 333 entries (95\%) met all quality criteria, demonstrating the effectiveness of our automated refinement approach.
Following this initial validation, we partition the refined dataset into training, validation, and test sets with sizes of 4,074, 300, and 300 samples, respectively, ensuring proportional representation of each security type across all subsets.

\noindent\textbf{Test Set Quality Control:} To establish a reliable evaluation benchmark, the same two experts conducted additional quality control specifically on the test samples (262 samples with security issues).
Through meticulous examination, they identified 83 samples requiring content clarification or enhancement, which were then collaboratively refined to ensure strict adherence to our secure code review criteria. This collaborative refinement process required experts to clarify technical descriptions, enhance impact analyses, and optimize remediation advice specificity. The whole quality control process took approximately 87.3 person hours.

\subsection{Secure-aware Fine-tuning}
While standard end-to-end instruction-based fine-tuning for review comment generation enables LLMs to produce feedback, this approach fails to effectively identify security issues or provide context-specific actionable suggestions, often resulting in inaccurate or overly generic comments.
To this end, we propose a new secure-aware fine-tuning strategy, which fine-tunes LLM to focus on generating code review comments capable of accurately identifying security issues and providing actionable fix suggestions, leveraging our curated dataset.
Specifically, we refine the training objective by modifying the loss function to prioritize two criteria: precise categorization of security issue types and heightened attention to security-critical code elements in code diffs. This approach enhances the model's capacity to produce context-sensitive, security-focused feedback, ultimately strengthening the efficacy of automated secure code reviews.

To achieve this, we introduce specific token sets that highlight security-critical elements within the code by adjusting the weighting scheme. These sets are defined as follows:

\begin{itemize}
\item $\mathcal{I}_V$: The set of tokens corresponding to identifiers in code changes referenced in review comments $R$ receive additional weighting, as these elements are critical for pinpointing security issues (\textit{e.g.}, insecure function usage, improper array indexing, \textit{etc}).
\item $\mathcal{I}_{ST}$: The set of tokens representing the specific security type (\textit{e.g.}, Input Validation) also receive additional weighting
\end{itemize}

Building upon this foundation, we design our secure-aware (SA) loss function $\mathcal{-L}_{SA}$ as follows:
\begin{align}
\mathcal{-L}_{SA} &=
\sum_{t \in R} \log P(x_t|x_{<t}) + \alpha \sum_{t \in \mathcal{I}_{V}} \log P(x_t|x_{<t}) \notag   \;+\\
&\quad  \beta \sum_{t \in \mathcal{I}_{ST}} \log P(x_t|x_{<t})
\end{align}
where \(x_i\) denotes the token at position \(i\),
and \(P(x_i \mid x_{<i})\) is the probability of generating \(x_i\) based on the proceeding tokens $x_{<i}$. 
The first part of the equation calculates the standard cross-entropy loss for all tokens in the review comment $R$. The second and third terms introduce a targeted upweighting for security-critical elements, \textit{i.e.}, tokens in \(\mathcal{I}_V\) and \(\mathcal{I}_{ST}\), 
modulated by coefficients \(\alpha\) and \(\beta\), respectively. This approach sharpens the model's focus on key security indicators. 

We adopted Low-Rank Adaptation (LoRA)~\cite{hu2021lora} to optimize our training process in a cost-effective manner. 

\subsection{Retrieval-augmented Review Generation}
To further improve the quality of generated review comments and mitigate the hallucination issues commonly encountered in LLMs, we leverage the RAG technique to incorporate specialized security domain knowledge.
RAG is a widely adopted paradigm that improves LLMs by integrating relevant information retrieved from external databases into the input~\cite{gao2023rag}, and has been widely used in various code-related tasks~\cite{shi2022race,zhang2023repocoder,wu2024repoformer}.
We first construct a template datastore consisting of high-quality code review comment templates, and then retrieve the most similar comment from the datastore based on the code under review and incorporate it into the generation process.

\noindent \textit{Template datastore construction.} 
Following established RAG practices~\cite{wang2022training} that build retrieval datastores from training data, we use our fine-tuning dataset to create templates. Given the current landscape of limited high quality secure code review data, this approach maximizes resource utilization.
We manually crafted 261 high-quality code review comment templates from the training set adhering to our previously defined structure for security code review comments ($R = (ST, D, I, A)$), encompassing all security types presented in Table~\ref{tab:final_categories}, with a distribution that closely approximates the proportional representation of each security type in the training dataset.
These templates serve as a knowledge base for generating high-quality review comments. 

\noindent \textit{Retrieval-Augmented Review Generation (RARG).}\label{RAG}
In this process, we employ a two-stage strategy. \ourmodel{} first generates an initial review comment based on the code change, from which we extract the corresponding security issue type. In the second stage, we utilize the BM25 algorithm to retrieve the most relevant review comment template. This retrieval process uses the code change as the query and the set of code changes linked to the predicted security issue type within the template library as the document corpus.

\begin{figure}[t]
    \centering
    \setlength{\abovecaptionskip}{0.1cm}
    \includegraphics[width=0.9\linewidth]{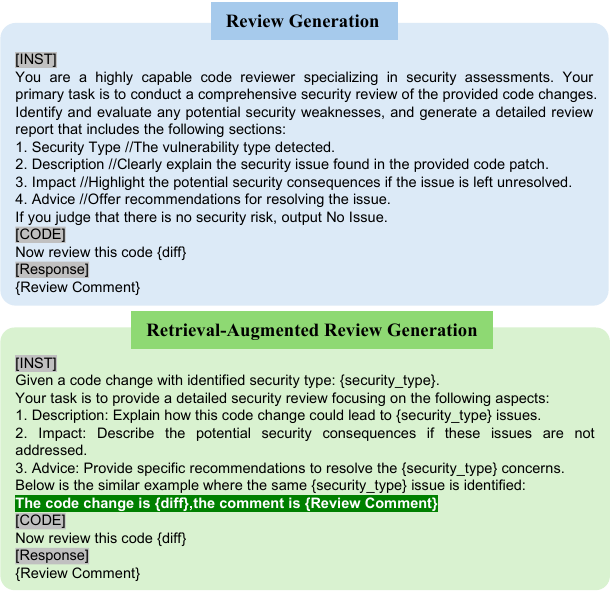}
    \caption{Prompt templates used for review generation.}
    \label{fig:template}
    \vspace{-0.4cm}
\end{figure}

The retrieved template serves as an auxiliary context of the prompt to guide the generation of the final review comment, ensuring it is more accurate and normative.
It is important to note that \textit{incorporating RAG does not affect the model's performance on issue detection since the retrieval is based on the predicted issue and does not alter the issue type within the review comment; instead, it solely updates other aspects of the comment's content.}
The prompt templates used in this process are illustrated in Figure~\ref{fig:template}.

\section{Experimental Setup}

\subsection{Metrics}
\subsubsection{Issue Detection}
For this task, following existing work~\cite{yu2024fine,li2022codereviewer}, we employ conventional metrics, \textit{Precision}, \textit{Recall}, \textit{F1-score}, and \textit{Accuracy}, to quantitatively assess the model's capability to accurately identify specific security issue within the framework of an 8-category classification task (7 security types + 1 non-issue type). The security type is extracted from the generated comment ($ST$).

\subsubsection{Review Comment Generation} 
For this task, \textit{samples with a reference security type of ``Non-Issue'' are excluded from this evaluation, as no review comments are expected for these cases}, resulting in 262 samples from the test set being evaluated (38 out of the original 300 were excluded).
To evaluate the quality of generated secure review comments, we use both \textit{BLEU-4} score and a new metric we designed, \textit{\textbf{SecureBLEU}}. 
Given that BLEU score struggles to fully assess the effectiveness of review comments in detecting and resolving security issues, we design SecureBLEU to capture both the general linguistic similarity between generated and reference texts and the critical inclusion of security-specific content. 

As shown in Algorithm~\ref{alg:SecCodeBLEU}, the metric computes a score by combining two components. The first component is a modified BLEU score ($\text{score}_{\textit{bleu}}$) evaluated across multiple fields of the review comment: security type, description, impact, and advice. The security type field is assessed through direct comparison—yielding a score of 100 for an exact match and 0 otherwise—while the remaining fields (description, impact, and advice) are evaluated using BLEU-4. 
The second component ($\text{score}_{\textit{keywords}}$) evaluates the overlap of security-specific keywords (associated with the detected security type $ST$) within the description, impact, and advice. These keywords are identified using a predefined dictionary $K[ST]$, where $K$ is a keyword dictionary organized by security type. 
The final score for each instance is calculated by equally weighting these two components, which was empirically validated in Section~\ref{bleu_weight}, ensuring a balanced assessment of both linguistic quality and security relevance while maintaining alignment with human judgment. $\text{Score}_{\textit{bleu}}$ assesses overall linguistic similarity, treating security-critical keywords no differently than ordinary words. In contrast, $\text{score}_{\textit{keywords}}$ specifically targets these security-critical terms as independent indicators of technical accuracy and domain expertise.

\subsubsection{Rationale behind SecureBLEU}
To justify the rationale behind SecureBLEU, we provide a detailed breakdown of how its two components work complementarily in Figure~\ref{fig:bleu_exp_new}, where the model \textbf{incorrectly classified a security issue from ``Access Control and Information Security'' to ``Type and Data Handling''}. Traditional BLEU-4 scores this comment at 26.44, focusing primarily on surface-level linguistic similarity. However, SecureBLEU's two-component analysis reveals critical deficiencies: (1) $\text{score}_{\textit{bleu}}$ = 17.57, where the incorrect security type classification (ST field = 0) penalized the overall linguistic assessment. This penalty mechanism is implemented through our modified BLEU computation in Algorithm~\ref{alg:SecCodeBLEU}, which incorporates the security type accuracy multiplier to ensure that misclassified security types receive substantially reduced scores regardless of surface-level text similarity. (2) $\text{score}_{\textit{keywords}}$ = 6.67, indicating that the model fails to include critical security keywords like ``unauthorized access'', ``misconfiguration'', and ``environmental security'' in its generated review comment. This omission occurred primarily due to misclassification of the security issue type, significantly impairing the comment's ability to properly address the ``Access Control and Information Security'' concern.
The final SecureBLEU score of 12.12 through weighted averaging provides a more justifiable quality assessment that BLEU's surface-level matching failed to capture, demonstrating how the two components together expose both linguistic and technical inadequacies in security-focused code review.

\subsection{Baselines}
We select a diverse set of baselines, including both specialized code review models and general-purpose LLMs, to ensure a comprehensive comparison with our proposed method. 

\begin{itemize}
    \item \textbf{CodeReviewer}~\cite{li2022codereviewer}: A pre-trained model specifically designed for code review. We fine-tuned the model on our dataset using the official code scripts and recommended hyperparameters.
    \item \textbf{LlamaReviewer}~\cite{lu2023llama}: A fine-tuned LLaMA model for code review tasks. We fine-tuned the model on our dataset, maintaining the same LoRA configurations as in our experiments.
   \item \textbf{General LLMs}: We evaluated a wide range of general-purpose LLMs that have shown strong performance in code-related tasks. These models span diverse architectures and parameter scales, including \textbf{GPT-4o}~\cite{achiam2023gpt4}, \textbf{Claude-3.5-sonnet}~\cite{claude}, \textbf{DeepSeek-V3}~\cite{liu2024DeepSeekv3}, \textbf{DeepSeek-R1}~\cite{guo2025DeepSeekr1}, \textbf{DeepSeek-Coder-6.7B-Instruct}~\cite{guo2024DeepSeek}, \textbf{Codellama-7B-Instruct}~\cite{roziere2023code}, and \textbf{Qwen2.5-Coder-7B}~\cite{hui2024qwen2}. 
\end{itemize}

\subsection{Implementation Details}
Given their strong performance in code-related tasks and our $\sim$4K fine-tuning samples, we selected CodeLlama-7B~\cite{roziere2023code}, DeepSeek-Coder-6.7B~\cite{guo2024DeepSeek}, and Qwen2.5-Coder-7B~\cite{hui2024qwen2} as our backbone models. Their 6-7B parameter size offers an optimal balance of capacity and efficiency, mitigating overfitting risks.
We configured LoRA with parameters of $r=8$, \texttt{lora\_alpha}=16, and \texttt{lora\_dropout}=0.05. Training was conducted with a maximum token length of 2048, a batch size of 4, gradient accumulation of 8, and a learning rate of 3e-4. To ensure reproducible outputs, inference utilized deterministic generation with greedy decoding.
For baseline reproduction, we ensured fair comparisons by adhering to original specifications. CodeReviewer~\cite{li2022codereviewer} was fine-tuned using its official repository and recommended hyper-parameters. To isolate architectural differences, LlamaReviewer~\cite{lu2023llama} was adapted using identical LoRA configurations as our method. The hyper-parameters for the baseline versions of CodeLlama-7B, DeepSeek-Coder-6.7B, and Qwen2.5-Coder-7B were also kept consistent with our model's setup.

For the remaining baseline models—\textit{i.e.}, GPT-4o, Claude-3.5-Sonnet, and DeepSeek-V3/R1—we used API with consistent parameters: \verb|temperature=0.7|, \verb|top_p=0.7|, and \verb|frequency_penalty=0.5| to ensure fair comparison. To account for the stochastic nature of these models with temperature=0.7, we conducted three independent runs for each API-based model and report the mean and standard deviation results (in Table~\ref{tab:overall_results}).
For our SA loss function, after extensive experiments to balance the trade-off between generating fluent review comments and focusing on security-critical elements, we set the coefficients to $\alpha=2$ and $\beta=5$.
To ensure a fair evaluation and to eliminate potential output format bias, all models were trained/prompted to generate reviews in standardized format according to our definition in Section~\ref{refinement}.

\begin{algorithm}[t]
\caption{SecureBLEU}
\setlength{\belowcaptionskip}{0.1cm}
\footnotesize{
\label{alg:SecCodeBLEU}
\begin{algorithmic}[1]
\State \textbf{Input:} $R^p$: predicted review, $R^r$: reference review, $K$: security-specific keywords dict, W: weight dict for fields
\State \textbf{Output:} SecureBLEU score
\If{$R^p$[ST] = "Non-Issue"} \Return 0 \EndIf
\State $\text{score}_{\textit{bleu}} \gets 0$
\For{each field in \{ST, D, I, A\}} // First Term
  \If{field = ST}
    \State score $\gets 100$ if $R^p[field]$=$R^r[field]$, $0$ otherwise
  \Else
    \State score $\gets$ BLEU-4$(R^p[\text{field}], R^r[\text{field}])$
  \EndIf
  \State $\text{score}_{\textit{bleu}} \gets \text{score}_{\textit{bleu}}$ + score*W[field]
\EndFor
\State $\text{score}_{\textit{keywords}} \gets 0$ 
\For{each field in \{D, I, A\}} // Second Term
  \State $\text{keywords}^r$ $\gets$ extract\_keywords$(R^r[\text{field}], K[ST])$
  \State $\text{keywords}^p$ $\gets$ extract\_keywords$(R^p[\text{field}], \text{$\text{keywords}^r$})$
  \If{$|\text{keywords}^r| > 0$}
        \State $\text{ratio} \gets |\text{keywords}^p|/|\text{keywords}^r|$ 
    \Else
        \State $\text{ratio} \gets 0$ 
    \EndIf
  \State $\text{score}_{\textit{keywords}} \gets \text{score}_{\textit{keywords}}$ + ratio*W[field]
\EndFor

\State \Return $0.5*\text{score}_{bleu} + 0.5*\text{score}_{keywords}$
\end{algorithmic}
}

\end{algorithm}

\subsection{Dataset Construction Cost}
Our dataset construction relies on both GPT-4o and expert validation, incurring financial and human costs. For the LLM judge process in data collection, the LLM processed 13,611 candidate samples (10,840 from keyword matching and 2,771 from embedding matching), requiring an average of 230.43 input tokens per judgment.
For data refinement procedure, LLM handled 4,089 samples, consuming an average of 692.14 input and 193.73 output tokens per sample. The total dataset construction cost was approximately \$46 based on GPT-4o pricing (\$5/1M input, \$20/1M output tokens).
For the expert validation, each review took an average of 8–10 minutes per sample, totaling 98 person hours for the initial quality assessment (351 samples) and 87.3 person hours for test set refinement (262 samples, including collaborative enhancements).

\section{Experimental Results and Analysis}

To assess the effectiveness of \ourmodel{}, we conduct experiments to address the following research questions:

\begin{itemize}
    \item \textbf{RQ1: Overall Performance} - How does \ourmodel{} perform compared to state-of-the-art code review models in terms of (1) accuracy in security issue detection, and (2) overall quality of generated review comments?
    \item \textbf{RQ2: Ablation Study} - What is the contribution of each component in \ourmodel{} to its overall performance?
    \item \textbf{RQ3: Quality Analysis} - How effectively does \ourmodel{} address different types of security issues?
\end{itemize}

\begin{table*}[t]
    \setlength{\abovecaptionskip}{0.1cm}
    \centering
    \small
    \caption{Results on issue detection and review comment generation. For general LLMs, we conducted three independent runs and report the mean and standard deviation of the results.}
    \begin{tabular}{lcccc|cc}
    \toprule
     \multirow{2}{*}{\textbf{Model}}   & \multicolumn{4}{c|}{\textbf{Issue Detection}} & \multicolumn{2}{c}{\textbf{Comment Generation}} \\
         &  Precision & Recall & F1 & Accuracy & BLEU & SecureBLEU \\
    \midrule
    CodeReviewer             & 65.88 & 57.44 & 59.03 & 58.53 & 8.66  & 21.31 \\
    LlamaReviewer            & \underline{66.06} & \underline{60.29} & \underline{61.46} & \underline{61.20} & 9.20  & \underline{24.56} \\
    \midrule
    DeepSeek-R1             & 54.73 (±1.77) & 46.54 (±1.45) & 46.24 (±1.31) & 46.27 (±2.05) & 5.81 (±0.38)   & 15.84 (±1.27) \\
    DeepSeek-V3            & 62.83 (±0.29) & 51.89 (±0.38) & 53.31 (±0.43) & 53.56 (±0.51) & \underline{10.80 (±0.34) } & 21.84 (±0.92) \\
    DeepSeek-Coder-6.7B       & 36.30  &  18.55 &   15.58 & 23.23 & 6.85  & 16.00 \\
    CodeLlama-7B             & 14.69 & 12.35 & 6.22 & 17.39 & 4.26  & 11.68 \\
    Qwen2.5-Coder-7B    & 46.31 & 39.61 & 38.57 & 45.00 & 7.04  & 20.63 \\
    GPT-4o             & 58.74 (±0.52) & 52.66 (±043) & 53.18 (±0.35) &  54.50 (±0.44) & 7.60 (±0.26)  & 19.33 (±0.52) \\
    Claude-3.5-sonnet       & 60.74 (±0.46) & 53.56 (±0.51) & 52.81 (±0.65) & 54.27 (±0.51) & 8.83 (±0.24)  & 19.54 (±0.83) \\
    \midrule
    \ourmodel{}$_{CL}$       & 73.28 & \textbf{71.48} & \textbf{71.98} & 71.91 & \textbf{11.34} & \textbf{29.31} \\
    \ourmodel{}$_{DS}$       &  72.25 & 71.23 & 71.62 & \textbf{72.24} & 11.01 & 29.23 \\
    \ourmodel{}$_{QW}$  &  \textbf{73.56} & 70.64 & 71.60 & 71.33 & 9.35  & 28.76 \\
    \bottomrule     
    \end{tabular}
    \label{tab:overall_results}
    \vspace{-0.3cm}
\end{table*}

\subsection{RQ1: Overall Performance}\label{RQ1}

\subsubsection{RQ1-1: Performance of Issue Detection}

The left section of Table~\ref{tab:overall_results} presents the performance comparison on the issue detection task. 
Among all the baselines, LlamaReviewer and CodeReviewer, both fine-tuned on our constructed dataset, demonstrate superior performance compared to general-purpose LLMs. This highlights the effectiveness and importance of fine-tuning on domain-specific and high-quality datasets, enabling these models to outperform their general-purpose counterparts by leveraging domain-specific knowledge. 
\ourmodel{}, implemented in three variants (\ourmodel{}$_{CL}$ based on CodeLlama, \ourmodel{}$_{DS}$ based on DeepSeek-Coder, and \ourmodel{}$_{QW}$ based on Qwen2.5-Coder), consistently outperforms all baseline models across all evaluation metrics. This highlights the efficacy of our secure-aware fine-tuning strategy, which enhances the model's sensitivity to security-related classification tokens, thus achieving better results in identifying security issues. 
It is also worth noting that while CodeLlama initially struggles to detect security issues, its fine-tuned version, \ourmodel{}$_{CL}$, achieves significant performance improvements. This further underscores the effectiveness of our fine-tuning strategy in enhancing the model's capabilities.
Moreover, we observe that general LLMs such as Claude, DeepSeek-V3, Qwen2.5-Coder, and GPT-4o, while not fine-tuned, still achieve considerable performance, demonstrating their promising potential and performance in detecting security issues even without task-specific optimization. The consistent results across multiple runs further validate the reliability of these comparisons.

\subsubsection{RQ1-2: Performance of Review Comment Generation}
For the evaluation of review comment generation, we utilize both the BLEU-4 score and the SecureBLEU metric. 

While BLEU-4 measures general linguistic similarity, SecureBLEU provides a more nuanced assessment by focusing on security-specific content. The results are shown in the right portion of Table~\ref{tab:overall_results}.
Regarding BLEU-4 score, \ourmodel{}$_{CL}$ achieves a score of 11.34, outperforming the best-performing baseline (DeepSeek-V3) by 5\%. Among all the evaluated models, BLEU-4 scores show relatively modest variation, with most baselines falling between 7 and 10. However, DeepSeek-R1 and CodeLlama diverge significantly from this range, scoring 5.81 and 4.26, respectively. 
After analyzing the results, we observe that the lower performance of DeepSeek-R1 is primarily due to its overly divergent and unstructured reasoning processes during code analysis. Specifically, the model tends to engage in excessive and repetitive thinking patterns, frequently shifting between analytical approaches without fully developing any single line of reasoning. This leads to shallow analyses that overlook critical issues while emphasizing irrelevant aspects of the code. As for CodeLlama, it often fails to adhere to instruction guidelines, frequently repeating input code verbatim rather than providing meaningful feedback.

Unlike BLEU-4, the SecureBLEU metric, which measures the effectiveness of the review comment in detecting and resolving the security issues, reveals more pronounced differences across models.
This metric effectively captures variations in the quality of security-focused content within the generated comments, providing a more nuanced assessment of their relevance and utility in addressing security concerns.

Among the baseline models, LlamaReviewer achieves the highest SecureBLEU score of 24.56, aligning with its strong performance in issue detection. This correlation indicates that the quality of generated review comments is closely tied to the model's ability to detect security issues. In other words, if a model can accurately identify security vulnerabilities, it is more likely to produce clear issue descriptions, thorough impact analyses, and actionable remediation recommendations.
Among general-purpose LLMs, DeepSeek-V3 and Claude achieve promising performance, with SecureBLEU scores of 21.84 and 19.54, respectively, even exceeding that of the fine-tuned CodeReviewer (21.31), demonstrating the adaptability of these LLMs to security-related tasks despite their lack of domain-specific fine-tuning. Nevertheless, \ourmodel{} outperforms all baselines substantially, achieving a 19\% relative improvement over the best baseline, underscoring the effectiveness of our fine-tuning strategy combined with retrieval-augmented generation, which improves both the linguistic quality and security relevance of the generated comments.

\vspace{1mm}
\begin{custommdframed}
\textit{Answer to RQ1:} \ourmodel{} achieves state-of-the-art performance in secure code review. For issue detection, it achieves 17\% higher F1 score and 18\% better accuracy than the best-performing baseline. Regarding the quality of generated review comments, it exceeds the best baseline of 11\% in BLEU-4 and demonstrates approximately 19\% improvement in SecureBLEU.
\end{custommdframed}

\subsection{RQ2: Ablation Study}
We conduct an ablation study to evaluate each component's contribution in \ourmodel{}, including: (1) domain-specific fine-tuning to establish baseline security expertise, (2) security-aware loss optimization to enhance focus on critical security elements, and (3) retrieval-augmented generation to ground reviews in established security best practices.
We incrementally incorporate these components using CodeLlama-7B, DeepSeek-Coder-6.7B, and Qwen2.5-Coder-7B as backbone models. The results are presented in Table~\ref{tab:ablation_results}.

\begin{table}[h]
    \setlength{\abovecaptionskip}{0.1cm}
    \centering
    \scriptsize
    \caption{Ablation study results of \ourmodel{}.}
    \begin{tabular}{lcccc|cc}
    \toprule
     \multirow{2}{*}{\textbf{Model}}   & \multicolumn{4}{c|}{\textbf{Issue Detection}} & \multicolumn{2}{c}{\textbf{Comment Generation}} \\
         &  Precision & Recall & F1 & Accuracy & BLEU & SecureBLEU \\
    \midrule
    DeepSeek-Coder-6.7B  & 36.30  &  18.55 &   15.58 & 23.23 & 6.85  & 16.00\\
    + Fine-tuning & 70.70 & 68.76 & 68.90 & 68.23 & 11.08 & 26.27 \\
    + SA-Loss &72.25  &  71.23   & 71.62&72.24 & \textbf{11.27} & 28.79 \\
    + RARG (our model) & \textbf{72.25} & \textbf{71.23} & \textbf{71.62} & \textbf{72.24} & 11.01 & \textbf{29.23} \\
    \midrule
    CodeLlama-7B   &14.69   &  12.35   &  6.22   &17.39  &4.26 & 11.68\\
    + Fine-tuning  &71.64   &  69.95 &    71.09 &  70.23  &11.91 & 27.88\\
    + SA-Loss & 73.28   &   71.48    & 71.98&71.91& \textbf{12.46} &\textbf{29.69} \\
    + RARG (our model) & \textbf{73.28}   &   \textbf{71.48}     & \textbf{71.98} & \textbf{71.91}& 11.34& 29.31 \\
    \midrule
    Qwen2.5-Coder-7B & 46.31 & 39.61 & 38.57 & 45.00 & 7.04 & 20.63 \\
    + Fine-tuning & 71.12 & 68.42 & 68.84 & 68.67 & 9.41 & 27.61 \\
    + SA-Loss & 73.56 & 70.64 & 71.60 & 71.33& \textbf{9.47} & \textbf{29.21} \\
    + RARG (our model) & \textbf{73.56} & \textbf{70.64} & \textbf{71.60} & \textbf{71.33} & 9.35 & 28.76 \\
    \bottomrule     
    \end{tabular}
    \label{tab:ablation_results}
    \vspace{-0.3cm}
\end{table}

\begin{figure*}[t]
    \centering
    \setlength{\abovecaptionskip}{0.1cm}
    \begin{subfigure}[b]{0.3\linewidth}
        \includegraphics[width=\textwidth]{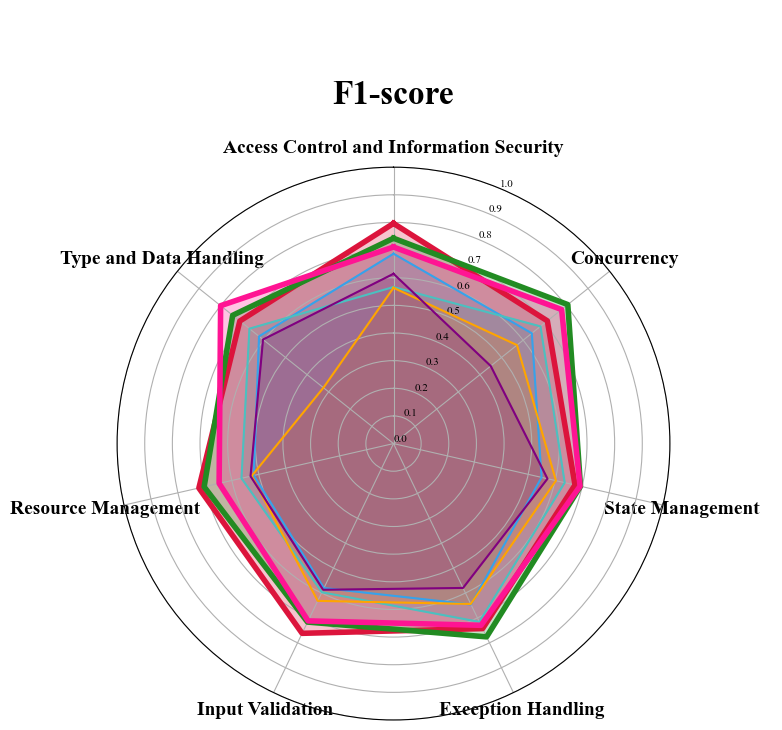}
        \caption{F1 score on issue detection.}
        \label{fig:f1}
    \end{subfigure}
    \begin{subfigure}[b]{0.3\linewidth}
        \includegraphics[width=\textwidth]{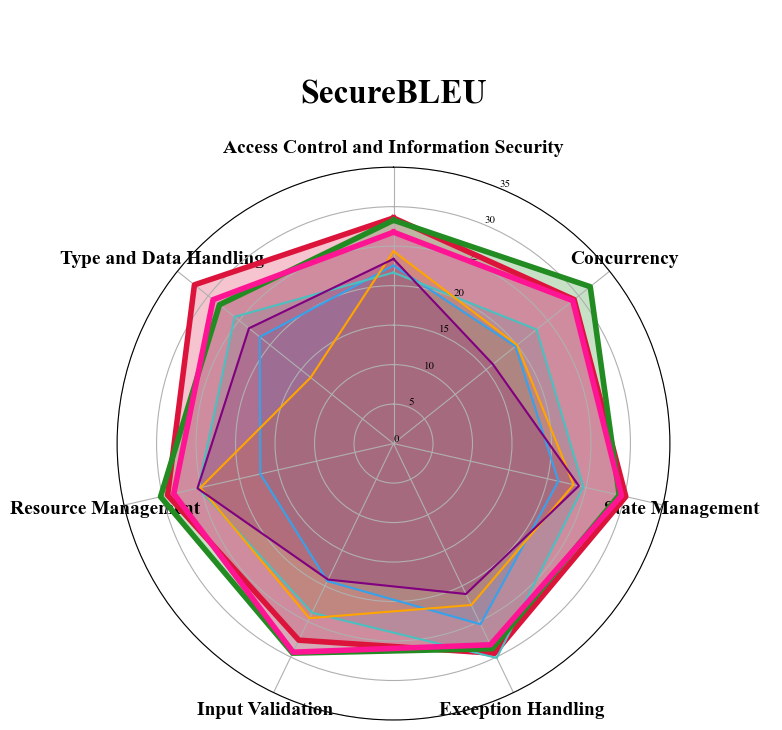}
        \caption{SecureBLEU score on review generation.}
        \label{fig:sb}
    \end{subfigure}
    \begin{subfigure}[b]{0.3\linewidth}
        \includegraphics[width=\textwidth]{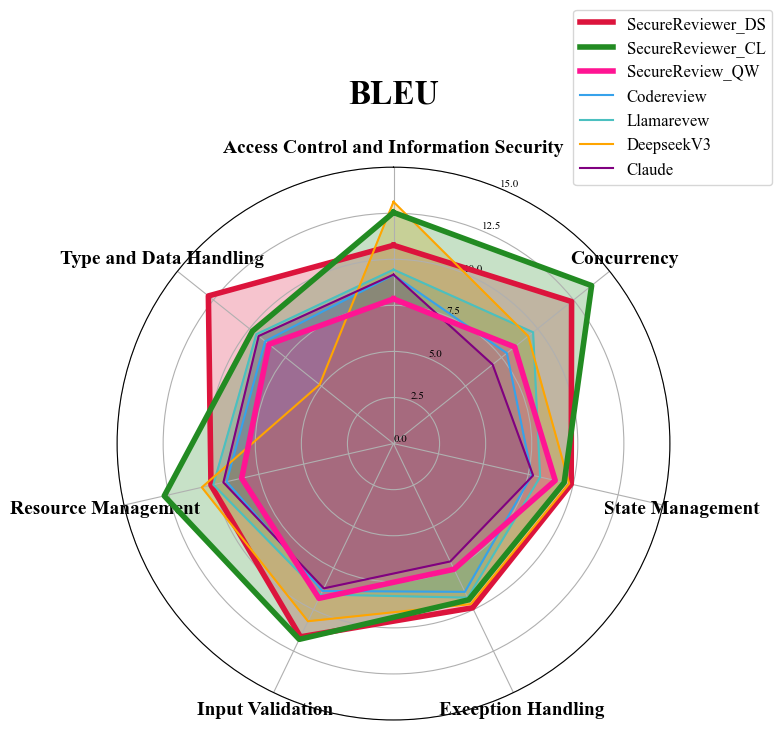}
        \caption{BLEU score on review generation.}
        \label{fig:bleu}
    \end{subfigure}
    \caption{Performance across various security types.}
    \label{fig:combined_scores}
    \vspace{-0.4cm}
\end{figure*}

\subsubsection{Impact of Fine-tuning} Domain-specific fine-tuning represents a fundamental adaptation strategy for LLMs to specialize in security-focused code review. We evaluate its contribution to enhancing our model's performance.

As seen from the results, the performance of vanilla LLMs reveals limited capability in secure code review. This is particularly pronounced in CodeLlama-7B, where poor instruction-following behavior significantly impairs its effectiveness. It frequently generates irrelevant identifiers, repetitive code snippets, and meaningless outputs, which may due to its insufficient exposure to security review tasks and code-diff patterns during pre-training.

Applying domain-specific fine-tuning yields substantial improvements, aligning the models with intricate code patterns and security vulnerabilities. For instance, fine-tuning CodeLlama-7B achieves an absolute improvement of 64.87 F1-score in issue detection, while Qwen2.5-Coder-7B improves by 30.27 F1-score, enhancing review comment quality with BLEU-4 increasing by 7.65 and SecureBLEU improving by 16.2 for CodeLlama-7B. These gains highlight fine-tuning's critical role in adapting general-purpose LLMs to the nuanced requirements of secure code review.

\subsubsection{Impact of SA-Loss}
After employing our proposed secure-aware loss optimization, which enhances the model's focus on security-critical tokens through a re-weighted loss function, the performance of both issue detection and review comment generation is further improved. Although the overall performance improvements are less pronounced compared to those achieved through domain-specific fine-tuning, this approach sharpens the model's sensitivity to security-related features, thus striking a better balance between precision and recall in issue detection and further enhancing the overall quality of the generated review comments.

\subsubsection{Impact of RARG} As mentioned in Section~\ref{RAG}, applying the RARG does not affect model's issue detection performance. As a result, the results of issue detection remain consistent with those achieved through fine-tuning and SA loss optimization. Regarding the review comment generation, we observe that applying the RARG does not yield consistent or significant improvements. 

This primarily stems from the following two aspects.
First, domain-specific capabilities instilled during fine-tuning render retrieved templates largely redundant with the model's internal knowledge base. The fine-tuned model already encodes specialized patterns for security issue detection and resolution, reducing the added value of external templates.
Second, fine-tuning may diminish the model's general instruction-following capacity, constraining its ability to leverage RAG's external context effectively. This is compounded by an inconsistency in instruction formats between training and inference phases. During fine-tuning, the model learns to generate comments without example-based instructions, whereas during RAG inference, a retrieved template is injected into the input prompt. This structural mismatch disrupts the model's ability to generalize under the altered input format, leading to suboptimal adaptation and diminishing returns.

Building on the aforementioned analysis, we argue that our RARG framework may prove particularly advantageous for general-purpose LLMs. To validate this hypothesis, we apply RARG to GPT-4o, Claude-3.5-Sonnet, and DeepSeek-V3, with results summarized in Table~\ref{tab:rag_performance}. As demonstrated in the results, RARG brings substantially improvements in SecureBLEU scores for these models. This notable improvement stems from a key distinction: general-purpose LLMs are inherently trained on broad, diverse datasets without task-specific specialization, rendering them deficient in domain-specific security knowledge compared to fine-tuned counterparts. The RARG framework effectively bridges this critical gap by retrieving and integrating security-relevant contextual patterns that these models would otherwise fail to prioritize. This supplementation enables them to produce more security-relevant and actionable review comments.

\begin{table}[h]
    \centering
    \small
    \setlength{\abovecaptionskip}{0.1cm}
    \caption{Performance of RARG on general LLMs.}
    \begin{tabular}{lcc}
    \toprule
     \textbf{Model} & \textbf{BLEU} & \textbf{SecureBLEU} \\
    \midrule
      GPT-4o  & 7.60 & 19.33 \\
      + RARG  & 7.47& 23.93  \\
      \midrule
      Claude-3.5-sonnet & 8.83  &  19.54\\
      + RARG  & 7.75 &   29.34 \\
      \midrule
      DeepSeek-V3 & 10.80    & 21.84\\
      + RARG & 10.19 &  25.64 \\
    \bottomrule
    \end{tabular}
    \label{tab:rag_performance}
    \vspace{-0.3cm}
\end{table}

\vspace{1mm}
\begin{custommdframed}
\textit{Answer to RQ2:} Each component contributes to \ourmodel{}'s performance gains, with domain-specific fine-tuning delivering the most substantial improvements. While RARG provides limited benefits for fine-tuned models, it substantially enhances general-purpose LLMs by augmenting their security knowledge.
\end{custommdframed}
\vspace{0mm}

\subsection{RQ3: Quality Analysis} 

We analyze \ourmodel{}'s performance in issue detection and review generation across the seven security types detailed in Table~\ref{tab:final_categories}. We compare our model against four top-performing baselines—CodeReviewer, LlamaReviewer, DeepSeek-V3, and Claude-3.5-sonnet—with the results presented in Figures~\ref{fig:combined_scores}.

\subsubsection{Issue Detection}
Figure~\ref{fig:f1} illustrates the issue detection performance across various security types. We can observe that all three variants of \ourmodel{} achieve balanced performance in issue detection, outperforming baseline models across multiple categories. The results reveal that baseline performance degrades with higher vulnerability complexity. Specifically, \textit{Concurrency} issues---which require complex semantic reasoning about thread synchronization---show large performance gaps between baselines and \ourmodel{}. CodeReviewer and LlamaReviewer, fine-tuned with domain-specific data, display more balanced performance across all types. These findings underscore the importance of domain-specific fine-tuning for achieving robust and generalizable performance on security-focused tasks. Despite \ourmodel{}'s substantial improvements over baselines, its performance varies across different security types. The approach is less effective on \textit{State Management} and \textit{Resource Management} issues. Notably, while demonstrating significant gains for \textit{Concurrency} issues, they remain a particularly challenging category across all model variants. This variation is attributable to the fundamental differences in how these distinct issue types manifest.

Specifically, \textit{Concurrency}, \textit{State Management}, and \textit{Resource Management} issues require deeper semantic reasoning about thread synchronization, state transitions, and resource lifecycles that extend beyond isolated code diff contexts. These limitations highlight the tension between pattern recognition and comprehensive semantic reasoning in our approach, as further analyzed in Section~\ref{sec:error_analysis}.

Figures~\ref{fig:sb} and~\ref{fig:bleu} present the review comment performance across different security types, evaluated using SecureBLEU and BLEU metrics, respectively. For SecureBLEU, all variants of \ourmodel{} deliver balanced and superior performance across all categories, substantially outperforming the baselines. Notably, the score distribution of all models aligns closely with the F1-score trends observed in issue detection (Figure~\ref{fig:f1}). This consistency highlights a strong correlation between issue detection performance and the quality of generated review comments, as captured by SecureBLEU. 

Conversely, categories with lower F1 performance in issue detection like \textit{State Management} and \textit{Resource Management} show correspondingly modest SecureBLEU scores. On one hand, lower F1 scores indicate fewer successful predictions, resulting in reduced overlap of category-specific security keywords and consequently lower weighted SecureBLEU scores. On the other hand, human-crafted reference comments for \textit{State Management} and \textit{Resource Management} issues extensively incorporate contextual code identifiers and causal explanations (\textit{e.g.}, ``variable X not released leads to resource leak''), while our model still struggles to capture these contextual references.

Regarding BLEU scores, as shown in Figure~\ref{fig:bleu}, \ourmodel{} achieves comparable or slightly higher scores than baselines, though performance improvements are less pronounced compared to SecureBLEU and F1 gains. This discrepancy primarily arises from our proposed secure-aware fine-tuning strategy, which is specifically optimized to prioritize security-critical tokens over general linguistic fluency. By focusing on accurately detecting security issues and providing precise explanations and advice, \ourmodel{} generates comments that, while highly relevant to security, may differ from reference review comments in phrasing or structure, resulting in relatively lower BLEU scores despite enhanced practical utility for security review purposes.

\vspace{1mm}
\begin{custommdframed}

\textit{Answer to RQ3:} \ourmodel{} shows balanced and superior performance across various security types in identifying and addressing security issues. An obvious correlation exists between the issue detection accuracy and the quality of the generated review comments. However, issues requiring deep semantic understanding remain challenging due to limited context incorporation and the inherent capabilities of LLMs.
\end{custommdframed}

\section{Discussion}

\subsection{Human Evaluation}\label{human_eval}

Since automatic metrics do not always agree with the practical utility of the review, we conduct human evaluation to further assess the quality of review comments generated by \ourmodel{}.

\noindent\textbf{\textit{Procedure.}} We recruit two software engineers in the evaluation, each with over 6 years of experience in Java and Python development, code review practices, and expertise in CWEs. The core security concepts in CWE—such as injection attacks, access control flaws, and cryptographic issues—share common security principles across different languages, enabling our evaluators to assess review comments based on security concepts and impact analysis rather than language-specific syntax details.
These experts independently evaluated the 262 generated comments from the test set (cases labeled "Non-Issue" were excluded). For each data point, evaluators were presented with the code diff, its corresponding reference comment, and the generated comment. Each generated comment was rated against four criteria drawn from academic research on code review effectiveness~\cite{chen2025understanding} and industry standards for security-focused reviews~\cite{owasp2023risk, metridev2023guidelines}:\ding{172} \textbf{Clarity}: Whether the review comment clearly explain the root cause of the issue, and references specific code snippets or patterns. \ding{173} \textbf{Relevance}: Whether the review comment relevant to the code contexts and issues, and avoid irrelevant or overly generic content. \ding{174} \textbf{Comprehensiveness}: Whether the impact analysis thoroughly explain potential consequences. \ding{175} \textbf{Actionability}: Whether the improvement advice specific, feasible, and aligned with best practices.
All ratings are integers on a scale of 1 to 5, with higher scores indicating better performance.

\begin{figure}[t]
    \centering
    \setlength{\abovecaptionskip}{0cm}
    \includegraphics[width=0.8\linewidth]{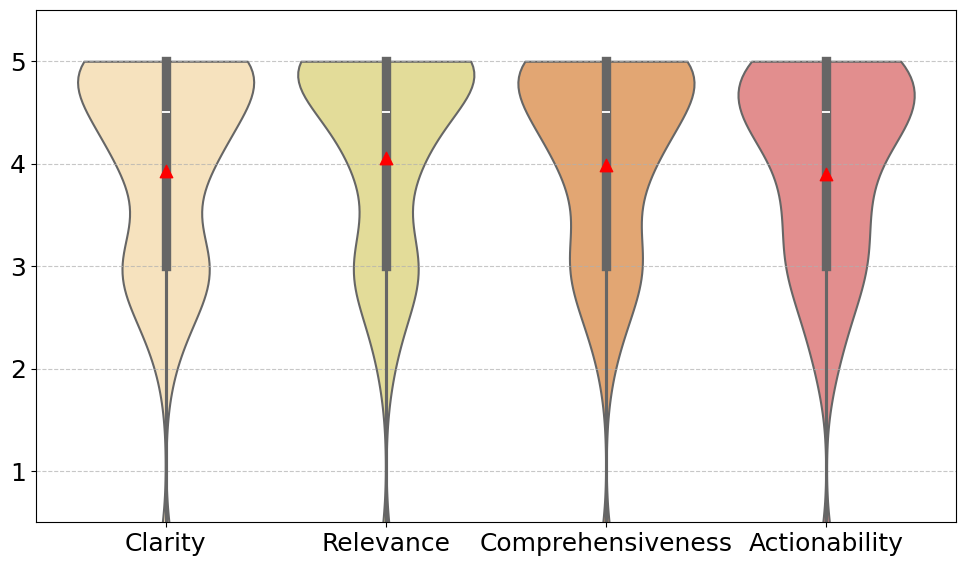}
    \caption{Human evaluation results of \ourmodel{}$_{DS}$.}
    \label{fig:human_eval}
    \vspace{-0.5cm}
\end{figure}

\noindent\textbf{\textit{Results.}} Figure~\ref{fig:human_eval} presents the results of the human evaluation.
Each score represents the average rating from two evaluators for the 262 test samples. 
A Cohen's Kappa coefficient~\cite{cohen1968weighted} of 0.66 confirms a substantial agreement between the raters.
The generated comments received consistently high ratings across all four criteria, with average scores of 3.93 for Clarity, 4.06 for Relevance, 3.98 for Comprehensiveness, and 3.90 for Actionability. These scores indicate that the reviews generated by \ourmodel{} demonstrate strong practical utility, effectively combining the proficiency to identify security issues with the ability to provide actionable guidance. This highlights the model's effectiveness in supporting real-world security review workflows.

\noindent\textbf{\textit{Correlation with SecureBLEU\&BLEU.}} We further calculate the Pearson's correlation between the human evaluation score with the two metrics used in our evaluation (BLEU and SecureBLEU).
The values are \( r=0.7533 \) and \( r=0.4026 \) for SecureBLEU and BLEU, respectively, which validate the strong alignment of SecureBLEU with human judgment.

\begin{figure}[t]
    \centering
    \setlength{\abovecaptionskip}{-0.4cm}
    \begin{subfigure}[b]{0.4\linewidth}
        \includegraphics[width=\linewidth]{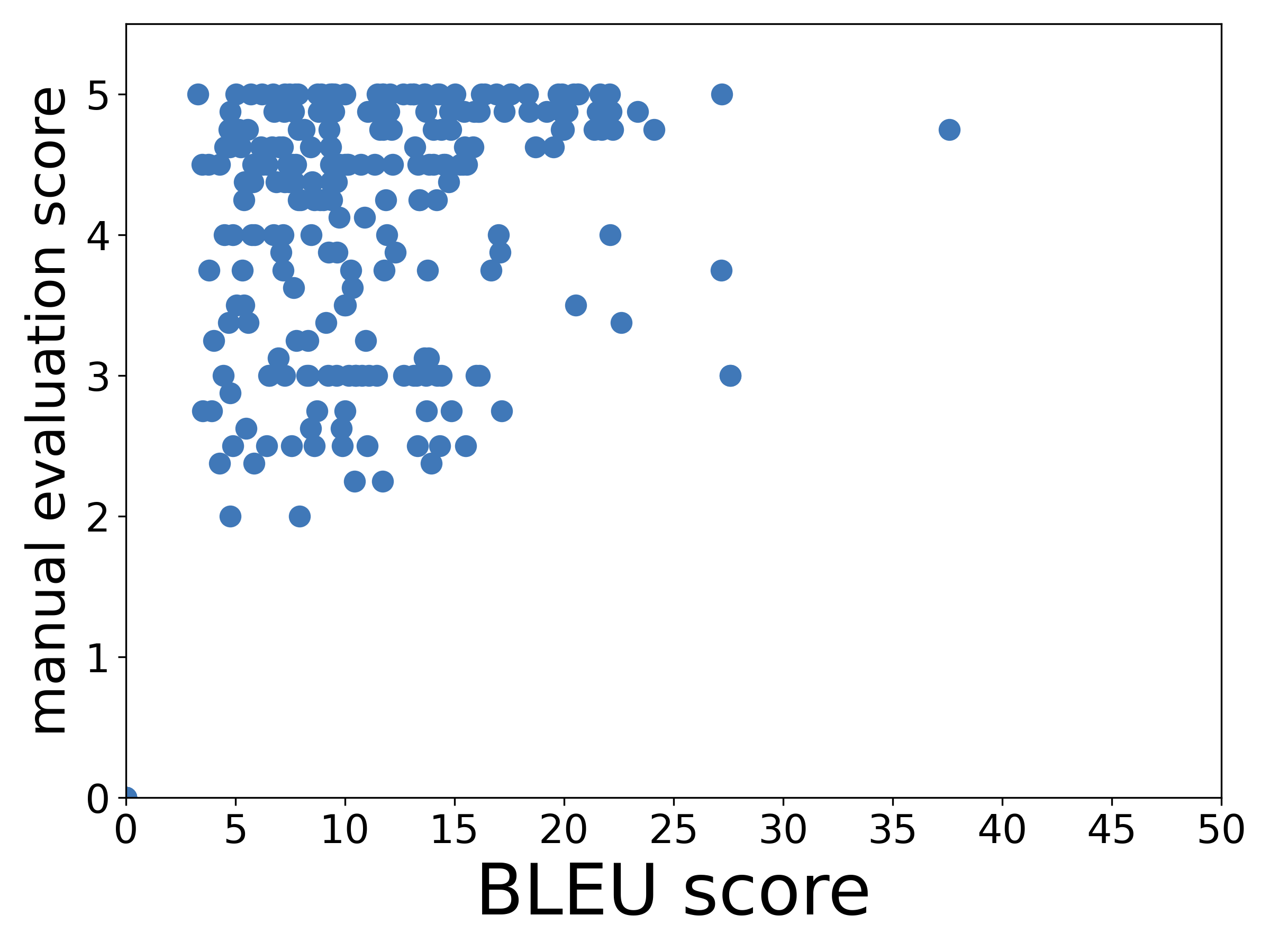}
        \label{fig:belu_human_eval}
    \end{subfigure}
    \begin{subfigure}[b]{0.4\linewidth}
        \includegraphics[width=\linewidth]{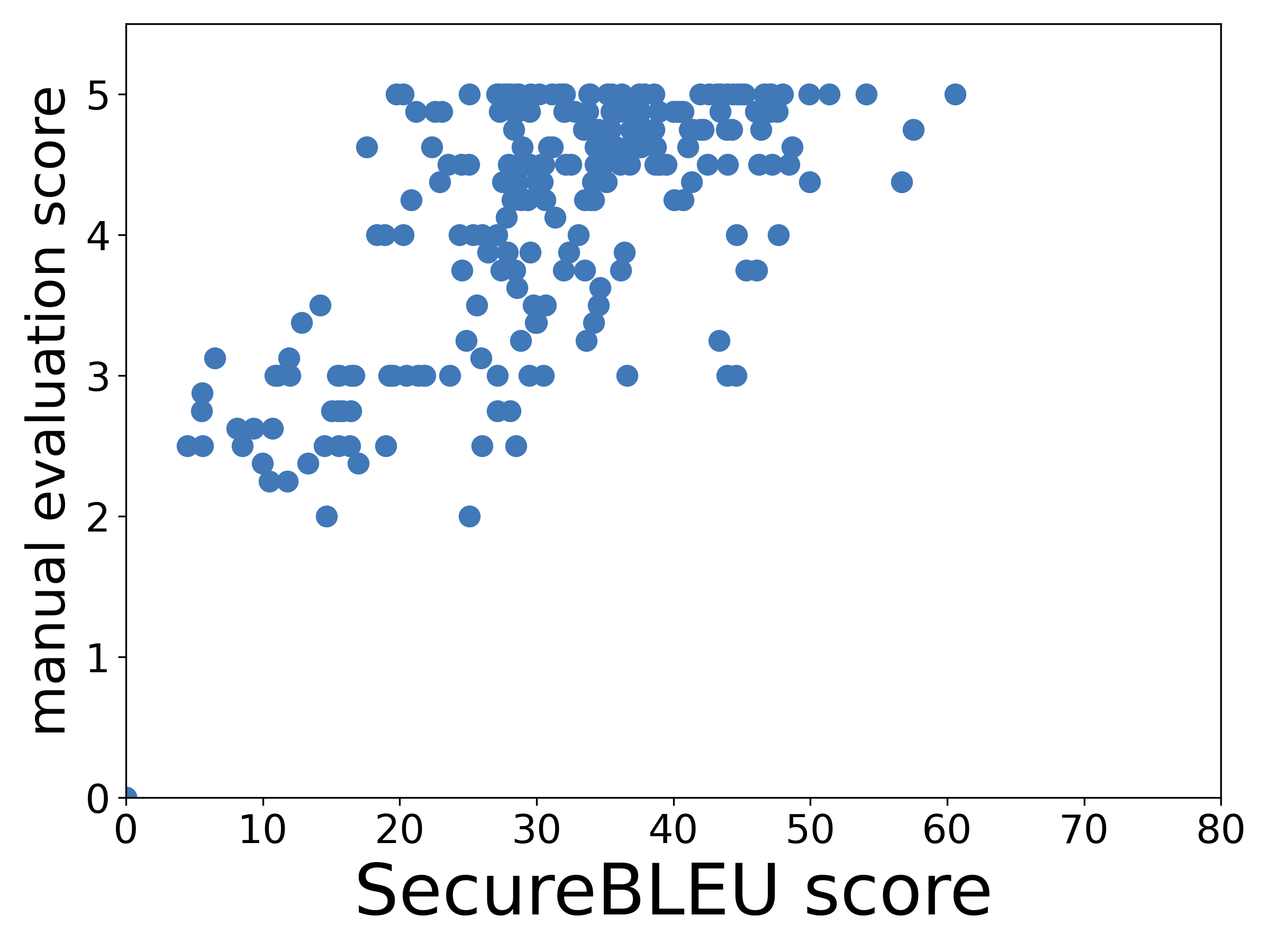}
        \label{fig:securebelu_human_eval}
    \end{subfigure}
    \caption{Correlation between human evaluation scores and SecureBLEU\&BLEU scores.}
    \label{fig:correlation}
    \vspace{-0.5cm}
\end{figure}

Figure~\ref{fig:correlation} shows the distribution of human evaluation scores alongside the corresponding BLEU and SecureBLEU scores. Notably, SecureBLEU exhibits a stronger correlation with human judgment compared to BLEU. The BLEU plot shows numerous points in the top-left region, where comments received low BLEU scores but high human ratings, indicating BLEU undervalues comments that human experts consider high quality.
In contrast, the SecureBLEU plot shows a more desirable distribution with fewer inconsistently evaluated points, broader score range, and stronger clustering of high human ratings with high SecureBLEU scores. 
These findings further confirm SecureBLEU's superior alignment with human preferences, establishing it as a more reliable metric for the automated evaluation of secure code review.

\noindent\textbf{\textit{Error Analysis.}}\label{sec:error_analysis}
To gain deeper insight into the limitations of \ourmodel{}, we perform a thorough error analysis of review comments that received low human evaluation scores and identify the following two error patterns.
\ding{182} \textit{Superficial Pattern Matching}: \ourmodel{} occasionally prioritizes surface-level pattern recognition-such as security-related keywords (\textit{e.g.}, map) or syntactic structures (\textit{e.g.}, mutex operations)-over in-depth semantic reasoning. For instance, when analyzing code with map operations, our model may focus on superficial indicators like the presence of map keywords or deletion operations. Thus, it erroneously flags concurrency issues (e.g., suggesting mutex protection for maps) but fails to diagnose the underlying root cause of the actual vulnerability, such as missing input validation or array bounds checking.

\ding{183}\textit{Limited Contextual Awareness}: In some cases, our model struggles to accurately interpret code semantics due to its reliance on isolated code diff, which lack the broader context of the full codebase, execution flows, and inter-procedural dependencies, resulting in failures on identifying possible issues. For example, when reviewing array operations, \ourmodel{} may fail to detect out-of-bounds vulnerabilities because it cannot infer how the array is initialized or modified in other parts of the codebase. 
These findings highlight the need for enhanced semantic understanding and broader contextual integration in automated code review, which will be the focus of our future work to further improve the effectiveness of our model.

\subsection{Impact of Weight Setting for SecureBLEU}\label{bleu_weight}
When computing SecureBLEU, we employ an equal weighting scheme for $\text{score}_{\textit{bleu}}$ and $\text{score}_{\textit{keywords}}$ to balance linguistic quality and security relevance. To validate this choice, we empirically compared different weighting schemes by measuring Pearson correlation between human evaluation scores and SecureBLEU (following Section~\ref{human_eval}). We systematically evaluated weight ratios for ($\text{score}_{\textit{bleu}}$, $\text{score}_{\textit{keywords}}$) ranging from 0.2/0.8 to 0.8/0.2. The results demonstrate that the 0.5/0.5 setting achieves the highest correlation coefficient (r = 0.7533) with human preferences, outperforming all alternative configurations. This confirms that an equal weighting aligns best with human judgment, thereby justifying our choice for computing SecureBLEU.

\subsection{Threats to Validity}
\textbf{Threats to internal validity} relate to the hyper-parameters setting during the fine-tuning. For API-based models, we used temperature=0.7 and conducted three runs to ensure statistical reliability. We conduct a small-range grid search on learning rate, batch size, LoRA parameters, the coefficients \(\alpha\) and \(\beta\) in SA loss, and the final setting was selected based on the best performance observed on the validation set. It is expected that more hyper-parameter tuning would bring more improvements.
Our study was also constrained by several factors. The limited size of our dataset restricted training to 7B backbone LLMs, and it is possible that newer models not included in our evaluation may offer superior performance. Furthermore, our use of the fine-tuning dataset to construct RAG templates, a decision necessitated by the scarcity of high-quality secure code review data, may limit the technique's effectiveness. Future work will explore additional LLMs and alternative datastores for RAG template construction.

\noindent\textbf{Threats to external validity} arise from potential errors in the LLM-based data refinement. We mitigated this by confirming that 95\% of the data met our quality standards via manual sampling (Section~\ref{refinement}) and by having two domain experts manually review and refine the entire test set. Moreover, our dataset construction relies on both LLM (GPT-4o) and expert annotation, incurring financial and human costs. Future work could leverage emerging cost-effective models like DeepSeek-V3 to reduce this expense while maintaining quality. While manual validation of the test set remains essential for ensuring reliable evaluation, this process can be optimized. Strategic automation, such as using CodeQL for initial security checks and leveraging cost-efficient LLMs for preliminary assessments and formatting, could further streamline this workflow.

\noindent\textbf{Threats to construct validity} relate to the rationality of evaluation metrics. Following existing code review research~\cite{li2022codereviewer,yu2024fine,lu2023llama}, we employ Precision, Recall, F1, and Accuracy to assess the model's capability to correctly identify security issues, and use BLEU-4 score and our proposed SecureBLEU to evaluate the quality of generated secure review comments. We further conduct human evaluation studies to assess the practical utility and quality of review comments generated by \ourmodel{}.

\section{Related Work}

\subsection{Code Review Automation}

Code review is a key practice in software development that involves a systematic review of the source code to find defects as well as improve quality. 
Recent advances in deep learning and LLMs have enabled significant progress in automating code review. 
~\citet{tufano2022using} fine-tune T5~\cite{raffel2020t5} model for generating review comments.~\citet{li2022codereviewer} introduce the CodeReviewer, a transformer-based model pre-trained with four pre-training tasks for code review. 
Recent advances using LLMs have shown promising results in both accuracy and interpretability.
~\citet{lu2023llama} propose the LLaMA-Reviewer using parametric efficient fine-tuning techniques to fine-tune the LLaMA model.~\citet{yu2024fine} fine-tune open-source LLMs with chain-of-thought-guided data to generate review comment that not only pinpoint code issues in detail but also provide logical explanations and actionable repair suggestions.
However, these general-purpose code review approaches often produce inaccurate or irrelevant comments, are impacted by dataset noise~\cite{tufano2024code,li2022codereviewer}, and lack security specialization. 
Moreover, current widely-adopted evaluation metrics, such as BLEU-4 score, also fail to address security-specific needs, underscoring the need for tailored automated techniques and evaluation frameworks focused on security issue detection.

\subsection{Code Review for Security Issues} 

Existing research on security-related code review predominantly focus on empirical studies, which investigate the role of code review in finding and mitigating security issues~\cite{yu2023security,charoenwet2023complementing,yu2024insight}. 

~\citet{charoenwet2023complementing} find that conventional reviews struggle with language-specific security issues, such as C++ memory management~\cite{lee2002study} or cross-site scripting attacks~\cite{hannousse2024twenty}. Similarly,~\citet{yu2023security} analyze 430,000 comments from open-source communities and reveal that security defects accounted for less than 1\% of discussions, with race conditions and resource leaks dominating the conversation. 
~\citet{charoenwet2024empirical} show that static analysis tools can detect certain vulnerabilities but falter when faced with context-dependent issues. Developers also resist these tools due to usability and integration difficulties, as noted by~\citet{johnson2013don}. More recently,~\citet{yu2024insight} demonstrate that LLMs outperform static analysis tools in detecting security defects but still face limitations in accuracy and contextual understanding. 

While these studies identify challenges and limitations in practice and provide valuable insights for improving secure code reviews, they fall short of providing automated solutions to systematically address these issues. 
These studies directly inspire our work in several key ways. For example, the security defect categories identified by~\citet{yu2023security} guide our selection of security-relevant keywords for the SecureBLEU metric and serve as filtering criteria during dataset construction.

\section{Conclusion}
In this paper, we propose \ourmodel{}, a framework designed for secure code review. We begin by constructing a dataset for training and evaluating the model's secure code review capabilities. Building on this foundation, we introduce a security-aware fine-tuning strategy to enhance the LLM's ability to generate code review comments that effectively identify security issues and provide actionable fix recommendations. Additionally, we integrate the RAG technique to mitigate LLM hallucinations and improve the relevance and reliability of generated comments. Experimental results demonstrate that \ourmodel{} outperforms state-of-the-art models, validating its effectiveness and practical applicability.

\begin{acks}
This research is supported by the National Natural Science Foundation of China Grants Nos. 62302021, 62332001, and the Fundamental Research Funds for the Central Universities (Grant No. JK2024-28).
\end{acks}

\bibliographystyle{ACM-Reference-Format}
\bibliography{ref}

\end{document}